
\magnification=1200
\tracingmacros=1
\font\ninerm=cmr9
\parindent=20pt\parskip=6pt plus 1pt
\baselineskip=15pt
\hsize=15.8truecm
\vsize=24.truecm
\def\finepagina{\vfill\eject}
\def\finerigo{\hfill\break}
\def\pmb#1{\setbox0=\hbox{#1}%
  \kern-.025em\copy0\kern-\wd0
  \kern.05em\copy0\kern-\wd0
  \kern-.025em\raise.0433em\box0 }
\def\bsigm{\pmb{$\sigma$}}
\def\calfam{cos~\alpha^{IM(-)}_{\mu\mu^{\prime}}}
\def\salfap{sin~\alpha^{IM(+)}_{\mu\mu^{\prime}}}
\def\delmu{\left({1+\delta_{\mu+\mu^{\prime},1}\over 1+\delta_{\mu,0}}
           \right)}
\def\dsigma{
    {d\sigma\over dE_{e^{\prime}}d\Omega_{e^{\prime}}d\Omega_{N}^{cm}}}
\def\ddsigma{
    {d\sigma/dE_{e^{\prime}}d\Omega_{e^{\prime}}d\Omega_{N}^{cm}}}

\def\fmumup{f_{\mu\mu^{\prime}IM}}
\def\fhmumup{f^h_{\mu\mu^{\prime}IM}}
\def\wmumup{w_{\mu\mu^{\prime}IM}}
\def\g#1{g^{#1}_{\mu\mu^{\prime} IM}}
\def\gh#1{g^{h #1}_{\mu\mu^{\prime} IM}}
\def\p#1{p^{#1}_{\mu\mu^{\prime} IM}}

\def\q2{q^2}
\def\qq2{{Q^2\over{\qlab^2}}}

\def\qcm{q_{cm}}
\def\qlab{q_{lab}}
\def\qlcm{{\qlab\over{\qcm}}}
\def\tmatrix{t_{s m_s \mu m_d}}
\def\tsmatrix{t^*_{s m_s \mu^{\prime} m^{\prime}_d}}
\def\ttsmatrix{t^*_{s^{\prime} m_s^{\prime} \mu^{\prime}
            m^{\prime}_d}}

\def\tmsq{t_{s m_s \mu m_d({\bf q})}}

\def\tg2{\tan^2{{\vartheta_{e'}}\over 2}}
\def\t2g{\tan^2(\vartheta_{e'}/2)}
\def\trejM{\pmatrix{1&1&I\cr m_d&-m^{\prime}_d&-M \cr}}
\def\trejpM{\pmatrix{1&1&I\cr m_d&-m^{\prime}_d&-M^{\prime} \cr}}
\def\vzero{v^{0}_{\mu\mu^{\prime}}}

\def\vl0{v^0_L}
\def\vt0{v^0_T}
\def\vtl0{v^0_{TL}}
\def\vtt0{v^0_{TT}}
\def\vht{v^h_T}
\def\vhtl{v^h_{TL}}
\def\flt{f_{LT}}
\def\fm-2{fm^{-2}}
\def\apz0y{P_{0y}}
\def\apu1x{P_{1x}}
\def\bpu1z{P_{1z}}
\def\apprd2x{P_{2x}^h}
\def\cpprd2z{P_{2z}^h}
\def\pz{{\bf P}_0}
\def\pu{{\bf P}_1}
\def\pd{{\bf P}_2}
\def\ppr{P^h}
\def\bppr{{\bf P}^h}
\def\ppru{{\bf P}_1^h}
\def\pprd{{\bf P}_2^h}
\def\pdeu{P^d_1}
\def\pded{P^d_2}

\nopagenumbers
\footline={\hss\tenrm\folio\hss} \pageno=1
\null
\vskip 2truecm
\centerline{\bf Nucleon polarization in deuteron electrodisintegration}
\bigskip\bigskip\smallskip
\centerline{B. Mosconi}
\centerline{\it Dipartimento di Fisica, Universit\`a di Firenze,
                I-50125 Firenze,Italy}
\centerline{\it Istituto Nazionale di Fisica Nucleare, Sezione di
                Firenze, I-50125 Firenze, Italy}
\bigskip
\centerline{J. Pauschenwein}
\centerline{\it Institut f\"ur Theoretische Physik, Universit\"at Graz,
                A-8010 Graz, Austria}
\bigskip
\centerline{P. Ricci}
\centerline{\it Istituto Nazionale di Fisica Nucleare, Sezione di
                Firenze, I-50125 Firenze, Italy}
\bigskip\bigskip\smallskip
{\ninerm
Outgoing nucleon polarization in exclusive deuteron electrodisintegration
is studied
at the quasi-elastic peak in the standard theory with emphasis on
the effect of nucleonic and pionic relativistic corrections.
The cases of polarized beam or/and
polarized target are considered.
Sizeable relativistic
effects are pointed out in several polarization components.
The sensitivity of nucleon polarization to
the neutron charge form factor $G_E^n$ is discussed.
In particular, it is shown that
the longitudinal component of neutron polarization with
vector polarized deuterons is as sensitive to $G_E^n$
as the sideways beam polarization transfer.
}
\bigskip\smallskip
\finepagina
\centerline{\bf I. INTRODUCTION}
\vskip 15pt
In this paper we shall study the outgoing nucleon polarization
in exclusive deuteron electrodisintegration with beam and target
polarized, taking into account the relativistic corrections (RC)
to the standard theory in a perturbation expansion approach [1-6].
Such theory, which in addition to the
non relativistic impulse approximation (IA) considers the
effects of meson exchange (MEC) and isobar excitation (IC)
currents, has proved to be rather successful for the
electromagnetic (EM) interactions in few-body nuclei at low
and medium energy-momentum transfers [7]. However, consideration
of nucleonic and pionic RC turns out essential for explaining
some observables in elastic electron scattering as well as in
photo- and electro-disintegration.
A well known case concerns
deuteron photodisintegration whose differential cross section
is heavily affected by RC particularly in the forward direction,
as our group first pointed out [8] to resolve the long standing
discrepancy in the energy spectrum of the $0^\circ$ cross
section measured at Mainz in 1976 [9]. Obviously, this large
relativistic effect in forward and backward direction strongly
modifies the shape of the differential cross section [8~b].
Indeed, the recent data on the proton angular
distribution taken with quasi monochromatic photon beams or with
absolutely calibrated neutron beams in the time reversed reaction,
are in good agreement with the relativistic
predictions [10]. For an exhaustive survey of deuteron
photodisintegration we refer to the review paper by Arenh\"ovel and
Sanzone [11].
\par
A second and more recent example is that of the longitudinal-transverse
structure function $\flt$
and thus, of the $\phi$-asymmetry ($A_\phi$)
of the $d(e,{e^\prime}p)n$ cross section.
A sizeable relativistic effect in $\flt$ has been found
by two of us [12] in a
paper devoted to the influence of RC on the exclusive cross
section and has been recently verified
by van der Schaar et al. [13] in a three-fold separation
experiment at NIKHEF-K. We would like to note that our results
of $A_\Phi$ [14] are hardly distinguishible from those obtained
by Hummel and Tjon [15], as reported in Ref.[13] , in a complete
covariant model based on the Bethe-Salpeter equation.
This makes us confident in the reliability of our
perturbative calculations limited to terms of order (v/c)$^2$
in the energy and momentum transfer region of the present
accelerators. Of course, the situation may well change at
higher energies where higher order terms
become necessary, because the perturbative expansion in power of
$(p/m)$ seems to converge
badly [16].
\par
An even stronger relativistic effect on $f^h_{LT}$,
the fifth structure function arising for polarized electrons, which was
predicted in Ref.[12], remains to be experimentally confirmed.
Note that an experiment on $f^h_{LT}$ or equivalently of the electron
asymmetry of the cross section
requires an out-of-plane spectrometer
in addition to a polarized
electron beam.
\par
Coming back to the nucleon polarization , its
first measurement in exclusive deuteron break-up
has been carried out very recently
using the longitudinally polarized electron beam
of the MIT - Bates Lab. [17] and a liquid
deuterium target, but the results are still to be published.
Such experiment, prohibitive in the past because of the
low duty cycle of the previous generation of accelerators
in spite of the improvements in nucleon polarimeters,
has been made possible by the upgrading of the
accelerator
and by the advances in the polarized electron source technology.
Note that there has not yet been a measurement of
${\bf P}_0$,
the polarization with unpolarized beam and target. The reason for choosing
the beam polarization transfer
$\bppr_0$ over $\pz$ in the first measurement of neutron
polarization lies in the interest of the sideways component
of $\bppr_0$ for emitted neutron (${\ppr_{0x}}(n)$) as a source
of information about the
poorly known charge form factor  of the neutron ($G_E^n$) as first
suggested by Arnold et al. [18] and studied by several authors [19-22].
In particular, Arenh\"ovel and collaborators [21,22]
have shown that ${\ppr_{0x}}(n)$ is almost independent
of NN potential models and of meson exchange effects in the
parallel kinematics. Unlike previous papers on outgoing nucleon
polarization [18-24] we shall consider
in detail also the target polarization transfer, i.e.
the polarization induced by
polarized deuterons,
because the rapid progress in the production
of polarized deuteron targets makes it possible to
foresee their measurement in the near future.
For completeness, we shall also consider the beam-target
polarization transfer.
\par
Much effort for electron scattering from polarized targets
is concentrated in the internal gas target experiments [25]
which have the
advantages of purity of the targets and of rapid polarization
reversal while an acceptable luminosity is obtained
even with thin polarized gas targets thanks to the high current
of the beam.
Presently, the devices used for feeding polarized
deuteron targets are the atomic beam sources which allow to reach
vector and tensor polarization close to unity.
Unfortunately, several depolarizing
effects (cell wall bounces, electron beam magnetic field,...)
may considerably reduce the polarization degree in the storage cell.
This technique has been successfully employed at Novosibirsk
VEPP-3 storage ring for an
experiment on tensor analysing power in elastic
electron deuteron scattering [26] and also
for the first measurement of the
tensor asymmetries of the exclusive inelastic
cross section as reported in Ref.[27].
The major problem with the atomic beam sources is the low intensity of
polarized atoms injected into the storage cell.
More than an order of magnitude
can be gained with the laser driven sources when operating in a high
magnetic field so compensating the figure of merit
for the lower values of deuteron
polarization achievable [28]. In conclusion,
it seems reasonable to expect in the near future
accurate data also on target polarization transfer
with the advent of facilities with continuous wave electron beams.
\par
In general, the purpose of the study of spin observables is to
exploit the enhanced sensitivity of such non-averaged observables
to small components of the transition amplitude in order to test
specific points of the theory (potential models, meson exchange
or quark effects,...) or to obtain information about badly known
quantities. A very interesting example is $G_E^n$ which could be
determined via
neutron polarization experiments in quasi free scattering.
We have recalled
above the sideways component of the beam polarization transfer.
As we shall see, the longitudinal component of the
target polarization transfer in scattering off vector polarized deuterons
is as sensitive as $P^h_{0x}(n)$ to $G_E^n$ models.
\par
A problem not yet settled in the theory of the EM nuclear interactions
is the choice of the EM form factors in the nucleon charge and
current operators, essentially the choice of the Dirac $F_1$ or the Sachs
$G_E$ form factor as nucleon charge density.
This uncertainty worries either IA operators and MEC operators.
For long time and by many authors it has been argued that the same
form factor should appear in IA current and in the (longitudinal
part of) meson exchange currents so as to satisfy the continuity
equation. This constraint has been questioned by Gross and Riska [29]
(see also Refs.[30-31])
on the basis that the EM form factor is naturally different in different
EM processes. For example, the pionic current is governed by
the pion form factor
$F_\pi$ and the contact current by the axial form factor $F_A$. They have
shown that it is possible to fulfill gauge invariance
while remaining free to use arbitrary EM form factor in IA and MEC
operators.
This is achieved by adding to the standard EM vertices
appropriate terms which make them able to
satisfy the Ward - Takahashi identity, which is the
gauge invariance constraint at the operator level.
Note that these additional terms are purely
longitudinal and thus vanish for on-shell particles. Moreover,
they do not contribute to transition amplitudes
because the electron current is conserved.
These prescriptions for the off-shell
extension of the EM vertices (as well as for the inclusion
of the hadronic FF in the meson propagators) have been criticized [32]
for their arbitrariness and also because of
their limited validity. Indeed,
the off-mass-shell form factors
cannot depend only on $Q^2$ as assumed in Ref.[29].
Clearly, this proposal has given a theoretical justification also
to calculations which go with $G_E$ in IA and with $F_1$ in MEC.
{}From the numerical point of view, this mixed case exhausts
the possibilities for pion MEC because $F_\pi$ and $F_A$
are rather close to the isovector part of $F_1$
in the intermediate $Q^2$ range.
\par
Some light on the problem is shed from considering the Dirac and Sachs
relativistic forms of the $\gamma NN$-vertex. In fact, in the non
relativistic reduction one obtains in the two cases corrections
of order (v/c)$^2$ which to a large extent compensate for the
sizeable differences in IA operators.
For example, the Dirac ($\rho^D$) and the Sachs ($\rho^S$)
charge density operators in the two-component
spinor space read
$$\eqalign{\rho^D&=F_1-(F_1 + 2 F_2){{{q^2}+
i{\bf \sigma}\times{\bf P}\cdot{\bf q}}\over{8 M^2}}~~~,\cr
\rho^S&=G_E - G_E {{q^2}\over{8 M^2}} -i(2 G_M- G_E)
{{{\bf \sigma}\times{\bf P}\cdot{\bf q}}\over{8 M^2}}~~~,\cr}\eqno(1)$$
for energy transfer $q_0$ negligible with respect to three-momentum
transfer $q$. In this kinematic conditions the relation between
$G_E$ and $F_1$ becomes $G_E = {F_1} - {F_2} {{q^2}/{4{M^2}}}$
and then $\rho^D$ and $\rho^S$ differ only by terms of fourth
order. Clearly, the $F_2$ part of the Darwin-Foldy
term in $\rho^D$, a substantial part of RC in the Dirac
parametrization, is already included in IA using the Sachs
parametrization.
\par
Consequently, calculations including RC should be little dependent
on the choice of $G_E$ or $F_1$ for processes dominated by the
nucleonic degrees of freedom
as the deuteron electrodisintegration in
the quasi-elastic region.
This has been verified by explicit evaluation of
selected examples of structure functions and spin observables in Ref.[14]
and by Arenh\"ovel et al. in their systematic study of the form factors
in the inclusive process [33] and of the structure functions in the
exclusive process [34] (see also Ref.[35] where in addition the frame
dependence of the calculations has been considered).
In particular, since MEC play only a minor role at the quasi-elastic peak,
the possible choices ($F_1~,~G_E~,~F_{\pi}~,~F_A$) of the EM form factor
in MEC do not give results significantly different.
In conclusion, we shall mainly report results of calculations with the Dirac
current and just comment
on the differences with calculations with the Sachs form
because, even if not definite, there are arguments in favour
of the Dirac form of the $\gamma NN$-vertex.
It follows by minimal
substitution from the Dirac Hamiltonian and at least for structureless
nucleons, it fulfills the Ward-Takahashi identity. Also, it is consistent
with the use of the Dirac nucleon propagator
in the evaluation of MEC and of pionic RC. Moreover,
it does not seem possible
to formulate a chiral invariant effective Lagrangian predicting
the Sachs form of the $\gamma NN$-vertex [4].
\par
Finally, we shall not deal with the potential model dependence
of the polarizations. All the calculations are done with
the Paris potential [36]. The reason for this is that the dependence
on potential models is not substantial in the quasielastic region [22].
\par
In Sect.2, we shall derive general formulas for cross section and nucleon
polarization when both beam and target are polarized. Our results
for all the components of nucleon polarization in the coplanar kinematics
will be presented in Sect.3. Finally, we shall summarize our work in Sect.4.
\finepagina
\centerline{\bf II. FORMALISM}
\vskip 15pt
In this paper we shall concentrate on
the outgoing nucleon polarization
in exclusive deuteron electrodisintegration with both beam and target
polarized. We shall restrict our considerations to the case where the deuteron
target is polarized with axial symmetry with respect
to an orientation axis ${\bf d}$ (characterized by the angles
($\theta_d,\phi_d$) with respect to the virtual photon momentum ${\bf q}$)
so that its state of orientation
is defined by two parameters, the vector $P^d_1$
and tensor $P^d_2$ parameters.
Note that this corresponds to the present
experimental situation for polarized deuteron target with the
quantization axis defined by an external magnetic field.
We refer to the paper by Dmitrasinovic and Gross [37] for the general
case of deuterons anyway polarized.
\par
Arenh\"ovel et al. [22] have already given general formulas
for cross section and, with
further limitation of unpolarized target, for outgoing nucleon
polarization. In the following we shall use the notations of Ref.[22]
apart from some minor modifications.
\par
The polarization of the nucleon detected in coincidence with the
electron is
expressed in terms of the $T$-transition matrices and the virtual
photon and deuteron density matrices by
$$\eqalign{&\left(\dsigma\right){\bf P} =
          {\sigma_M\over{M_d}} \sum_{\mu\mu^{\prime}} v_{\mu\mu^{\prime}} \cr
          &\times \sum_{s m_s s^{\prime} m^{\prime}_s}
                  \sum_{m_d m^{\prime}_d}
          T_{s m_s \mu m_d({\bf q})}
          \rho^d_{m_d m^{\prime}_d}
          T^{*}_{s^{\prime}m^{\prime}_s\mu^{\prime}m^{\prime}_d({\bf q})}
          \langle s^{\prime} m_s^{\prime} \vert \bsigm
          \vert s m_s \rangle
          ~~~~~,\cr}\eqno(2)$$
where $\bsigm$ is the spin operator of the nucleon. For convenience
we have factorized out in Eq.(2) the Mott cross section
$\sigma_M$ and the  deuteron mass $M_d$.
The  transition matrices $T_{s m_s \mu m_d({\bf q})}$
in Eq.~(2)  are so denoted to recall that $m_d$ is the deuteron
spin projection on the momentum transfer ${\bf q}$. The final state is
characterized by spin $s$ and its projection $m_s$ on the relative
momentum ${\bf p}_{cm}$.
We assume, as usual,
relativistic electron energies so that the electron beam may
only have longitudinal polarization of degree $h$.
Therefore, $v_{\mu\mu^{\prime}}$ consists of two terms,
$v_{\mu\mu^{\prime}}=v^0_{\mu\mu^{\prime}}+h v^h_{\mu\mu^{\prime}}$ which
correspond to unpolarized  and polarized electrons, respectively.
They are  kinematical functions depending only
on the electron scattering variables. Because of
the symmetry relations
$$\eqalign{
       v_{\mu\mu^{\prime}}&=v_{\mu^{\prime}\mu} ~~~~~~~~~~,  \cr
   v^0_{-\mu-\mu^{\prime}}&=(-)^{\mu+\mu^{\prime}}v^0_{\mu\mu^{\prime}}
                                                             ~~~~~, \cr
   v^h_{-\mu-\mu^{\prime}}&=(-)^{\mu+\mu^{\prime}+1}v^h_{\mu\mu^{\prime}}
   ~~~~,\cr}\eqno(3)$$
all the possible components can be simply derived from
%
%
%
%
%
%
$$\eqalign{
     \vl0&={{\left(\qlcm\right)}^2}~\xi^2~~,\cr
     \vt0&=\eta+{1\over 2} \xi~~,\cr
     \vtl0&={1\over\sqrt 2} \left(\qlcm\right)\xi {\sqrt {\eta+\xi}}~~,\cr
     \vtt0&=-{1\over 2} \xi~~,\cr
     \vht&= \sqrt {\eta(\eta+\xi)}~~,\cr
     \vhtl&={1\over \sqrt 2} \left(\qlcm\right) \xi \sqrt{\eta}
                  ~~,\cr}\eqno(4)$$
where the indices $L,\  T,\  TL$ and $TT$ correspond to $(\mu \mu^{\prime}) =
(00),\ (11),\ (10)$ and $(1-1)$; $q_{lab}$ and $q_{cm}$ are the moduli
of ${\bf q}$ in the laboratory ($lab$) and final n-p
center-of-mass ($cm$) frames;
$\xi=Q^2/q^2_{lab}$~~ ,  $\eta=\t2g$ ~, $Q^2={\bf q}^2 - q_0^2$
being the four-momentum transfer squared
and $\vartheta_{e^\prime}$ the electron $lab$ scattering angle.
\par
Note that the definitions (4) of the $v's$ include the appropriate
factors of $(q_{lab}/q_{cm})$ which are necessary because we
calculate the
nuclear matrix elements in the $cm$ frame.
\par
The dependence of the $T$-transition matrices on the angle  $\phi$
between the reaction plane and the scattering plane
can be separated out, defining the reduced $t$-matrices
$$T_{s m_s \mu m_d({\bf q})}=e^{i(\mu+m_d)\phi} ~~~\tmsq
  ~~~~.\eqno(5)$$
Clearly, in order to evaluate the reduced $t$-matrix elements one
has to refer both initial and final states to a common quantization
axis. To this end we have to express the final state with respect
to ${\bf q}$ or the deuteron state with respect to ${\bf p}_{cm}$.
In the second case we must
rotate the deuteron state through the
angles $(0,-\vartheta_{cm},0)$, $\vartheta_{cm}$ being the
polar angle of  ${\bf p}_{cm}$ with respect to  ${\bf q}$.
Hence, the reduced transition matrices must be transformed according to
$$\tmsq = \sum_{m^\prime_d} t_{sm_s\mu m^{\prime}_d({\bf p}_{cm})}
          ~~d^1_{m^{\prime}_d m_d}(-\vartheta_{cm})
          ~~~~.\eqno(6)$$
We shall neglect in the following the specification of the quantization
axis when not necessary.
The $t$-matrix elements are expressed by means of the charge
and current matrix elements as:
$$\eqalign{\tmatrix=&-\sqrt{{p_{cm}E^p_{cm}E^d_{cm}\over 16 \pi^3}}
        e^{-i(\mu + m_d)\phi} \cr
        &\times \langle s m_s \vert \delta_{\mu0} \rho({\bf q})+
        \delta_{\vert\mu\vert 1} {\bf e}_\mu \cdot {\bf j}({\bf q})
        \vert m_d \rangle~~~~~,\cr}\eqno(7)$$
having denoted by ${\bf e}_\mu$ the photon polarization vector
of helicity $\mu$.
Owing to the factorization of $\sigma_M/M_d$ in Eq.~(2)
the reduced $t$-matrix is dimensionless as that introduced in
Ref.[12]. The only difference is in
the kinematic factors which maintain in
Eq.(7)  the relativistic expression. As for the final
$np$ state $\vert s m_s \rangle$ it is normalized so that it
becomes
$$\vert s m_s \rangle = e^{i{\bf p}_{cm}\cdot{\bf r}} \chi_{s m_s}
                      \eqno(8)$$
in plane wave (PW) approximation.
\par
As said above, the deuteron density matrix is diagonal with respect to
the symmetry axis ${\bf d}$. Thus, its expression in terms of the
orientation parameters  $P^d_I (P^d_0=1)$ is
$$\rho^d_{m_d m^{\prime}_d} =
          \sum_{I=0}^2 \sum_{M^{\prime}=-I}^{+I}
          \sqrt{\hat I\over 3}
          (-1)^{1-m_d} \trejM e^{-iM \phi_d}
          d^I_{M0}(\theta_d) P^d_I~~~~,\eqno(9)$$
($\hat I =2I+1$) in the frame  with  ${\bf q}$ as quantization axis taking
account of the rotation which tranforms $P^d_I$ from the
${\bf d}$-frame to the ${\bf q}$-frame.
\par
Note that the deuteron density matrix is not affected by the Lorentz
transformation along ${\bf q}$, ${\bf q}$ being the quantization axis
of the deuteron state.
\par
Using expressions (5)-(9)
we can write the polarization of the detected nucleon in the form:
$$\left(\dsigma\right){\bf P}=
       {\sigma_M\over{M_d}} \sum_{I=0}^2 P_I^d
       \left[{\bf p}_I +h {\bf p}_I^h\right]
       ~~~~.\eqno(10)$$
If we define the quantity
$$\alpha^{IM(\pm)}_{\mu\mu^{\prime}}=
  [(\mu-\mu^{\prime})\phi + M(\phi-\phi_d)] \pm (I-1){\pi\over 2}
  ~~~~,\eqno(11)$$
the cartesian components of the polarization vectors are given by
$$\eqalign{
  p^{x/z}_I&= \sum_{M=-I}^{+I} d^I_{M0}(\theta_d) \sum_{\mu =0}^1
                 \sum_{\mu^{\prime}=-\mu}^{+\mu} \calfam~ \vzero \g{x/z}~~,\cr
  p^{y}_I  &= \sum_{M=-I}^{+I} d^I_{M0}(\theta_d) \sum_{\mu =0}^1
                 \sum_{\mu^{\prime}=-\mu}^{+\mu} \salfap~ \vzero \g{y}~~,\cr
  p^{hx/z}_I&= \sum_{M=-I}^{+I} d^I_{M0}(\theta_d)
                  \sum_{\mu^{\prime} =0}^1
                  sin~\alpha^{IM(+)}_{1\mu^{\prime}}~
                  v^h_{1\mu^{\prime}}
                  g^{hx/z}_{1\mu^{\prime} IM}~~,\cr
  p^{hy}_I  &= \sum_{M=-I}^{+I} d^I_{M0}(\theta_d)
                  \sum_{\mu^{\prime} =0}^1
                  cos~\alpha^{IM(-)}_{1\mu^{\prime}}~
                  v^h_{1\mu^{\prime}}
                  g^{hy}_{1\mu^{\prime} IM}~~,\cr}
                  \eqno(12)$$
with respect to the right-handed frame which, according to the
Madison convention, has the z-axis along the nucleon momentum
${\bf p}_{cm}$ and the y-axis along ${\bf  q} \times {\bf p}_{cm}$.
\par
The polarization structure functions (PSF)
$\g{x/y/z}$ and $\gh{x/y/z}$ are defined by:
$$\eqalign{\g{x} &= - \sqrt{2} \delmu Im \left(i^{I}[\p{1} - \p{-1}]\right)
                  ~~~~, \cr
           \g{y}  &=  \sqrt{2} \delmu Im \left(i^{-I}[\p{1} + \p{-1}]\right)
                   ~~~~,\cr
           \g{z}  &=  2 \delmu Im \Bigl(i^{I}\p{0}\Bigr)~~~~,\cr
           \gh{x}  &=  \sqrt{2} \delmu Re \left(i^{-I}[\p{1} - \p{-1}]\right)
                    ~~~~,\cr
           \gh{y}  &=  \sqrt{2} \delmu Re \left(i^{I}[\p{1} + \p{-1}]\right)
                    ~~~~,\cr
           \gh{z}  &= - 2 \delmu Re \Bigl(i^{-I}\p{0}\Bigr)~~~~,\cr}
           \eqno(13)$$
in terms of the $p$-functions
$$\eqalign{\p{\lambda}
          &=\sqrt{\hat I\over 3} \sum_{M^{\prime}=-I}^{+I}
           \sum_{m_d m^{\prime}_d} (-1)^{1-m_d}
           d^I_{M^{\prime}M}(-\vartheta_{cm}) \trejpM \cr
          &\times \sum_{s m_s s^{\prime} m_s^{\prime}}
           \tmatrix \ttsmatrix
           \langle s^{\prime} m_s^{\prime} \vert \sigma_{\lambda}
           \vert s m_s \rangle
           ~~~~~,\cr}\eqno(14)$$
where $\sigma_{\lambda}$ is the $\lambda=0, \pm 1$ spherical
component of the spin operator.
\par
Clearly Eq.(12) exhibits polarization components in a very symmetric and
compact form which however suffers from a minor and a major drawback.
The minor one is that the dependence on the azimuthal angle $\phi$ is
not explicit. The major one is that these components are written in
terms of more PSF than necessary. In fact, the PSF's (13) are not
all independent and even not all different from zero because of the
symmetry relations fulfilled by the $p$-functions:
$$\eqalign{p^{\lambda~*}_{\mu\mu^{\prime}IM}
          &=(-1)^{\lambda+M} ~~p^{-\lambda}_{\mu^{\prime}\mu I-M}~~~~~~~~,\cr
           p^{\lambda}_{-\mu-\mu^{\prime}IM}
          &=(-1)^{1+I+M+\lambda+\mu+\mu^{\prime}}
           ~~p^{-\lambda}_{\mu\mu^{\prime}I-M}
          ~~~~. \cr} \eqno(15)$$
\par
The first relation in Eq.(15) comes directly from definition (14) while the
second one needs in addition the symmetry relation induced on the
$t$-matrix elements by parity conservation
$$t_{s-m_s-\mu-m_d} = (-1)^{1+s+m_s+\mu+m_d} ~~t_{s m_s \mu m_d}
  ~~~~.\eqno(16)$$
\par
Note that equalities (15) have already been  used to reduce the summation
over $\mu$ and $\mu^{\prime}$ to the four cases denoted $L, T, TL, TT$.
Furthermore, it is straightforward to deduce from
Eq.(15) that the non-interference PSF with $M \leq 0$ differ
at most for the sign from those PSF with $M \geq 0$.
Explicitly we have :
$$\eqalign{g^{x/z}_{\mu\mu,I-M} &= (-1)^{I-M+1} ~~g^{x/z}_{\mu\mu,IM} ~~,\cr
           g^{y}_{\mu\mu,I-M} &= (-1)^{I-M} ~~g^{y}_{\mu\mu,IM} ~~,\cr
           g^{h~x/z}_{\mu\mu,I-M} &= (-1)^{I-M} ~~g^{h~x/z}_{\mu\mu,IM} ~~,\cr
           g^{h~y}_{\mu\mu,I-M} &= (-1)^{I-M+1} ~~g^{h~y}_{\mu\mu,IM} ~~,\cr}
           \eqno(17)$$
for $\mu$=0,1.
Therefore, the independent PSF are 121
(40 with $M=0$ and 81 with $M \not= 0$).
The use of relations (16) and (17) allows to
partially reduce the summation
over $M$ in some but not in all the polarization components,
with the consequence that the
high symmetry of expression (12) is lost. For this reason we prefer not
to further handle these expressions.
\par
For completeness, we also report the cross section in a similarly
compact form, namely
$$\dsigma={\sigma_M\over{M_d}} \sum_{I=0}^2
            P_I^d \left(F_I + h F^h_I\right) ~~~~.\eqno(18)$$
The functions $F_I, F_I^h$ are given by
$$\eqalign{F_I&=\sum_{M=-I}^{+I} d^I_{M0}(\theta_d) \sum_{\mu =0}^1
               \sum_{\mu^{\prime}=-\mu}^{+\mu} ~\salfap \vzero \fmumup~~~~,\cr
         F_I^h&=\sum_{M=-I}^{+I} d^I_{M0}(\theta_d)
                \sum_{\mu^{\prime} =0}^1
                 cos \alpha^{IM(-)}_{1\mu^{\prime}}~
                 v^h_{1\mu^{\prime}}
                 f^{h}_{1\mu^{\prime} IM}~~~~,\cr}
                \eqno(19)$$
in terms of the structure functions (SF)
$$\eqalign{\fmumup  &= - 2 \delmu Re(i^{-I} \wmumup)~~~~,\cr
           \fhmumup &=  2 \delmu Im(i^I \wmumup)~~~~,\cr}
           \eqno(20)$$
where
$$\eqalign{\wmumup=&\sqrt{\hat I\over 3} \sum_{M^{\prime}=-I}^{+I}
           \sum_{m_d m^{\prime}_d} (-1)^{1-m_d}
           d^I_{M^{\prime}M}(-\vartheta_{cm}) \trejpM \cr
          &\times \sum_{s m_s}
           \tmatrix \tsmatrix  ~~~~.\cr}\eqno(21)$$
\par
Here again there are symmetry relations among the auxiliary
functions $\wmumup$ :
$$\eqalign{w^{*}_{\mu\mu^{\prime}IM}
    &=(-1)^{M} w_{\mu^{\prime}\mu I-M} ~~~~, \cr
           w_{-\mu-\mu^{\prime}IM}
    &=(-1)^{I-M+\mu+\mu^{\prime}}
           w_{\mu\mu^{\prime}I-M} ~~~~, \cr}
    \eqno(22)$$
which have already been used to write Eq.(19) in terms of only
$\mu=0,1 ; -\mu\leq\mu^{\prime}\leq\mu$.
\par
Analogously to the symmetry relations (15) for the $p$-functions,
the first equality (22) is an immediate consequence of definition (21),
while the second one follows from the parity conservation relation (16).
Owing to Eq.(22)
$$\eqalign{f_{\mu\mu,I-M}&=(-1)^{I-M}~~ f_{\mu\mu,IM}~~~~, \cr
           f^h_{\mu\mu,I-M}&=(-1)^{I-M+1}~~ f^h_{\mu\mu,IM}~~~~, \cr}
           \eqno(23)$$
and therefore the 54 SF appearing in Eq.(20) reduce to 41 independent SF
(14 with $M=0$ and 27 with $M \not= 0$).
The explicit expression of the cross section in terms of the independent SF
can be found in Ref.[34].
\par
The usual definition of the polarization variables is
$$\left(\dsigma\right){\bf P}=
       \left(\dsigma\right)_0 ~~\sum_{I=0}^2 P_I^d
       \left[{\bf P}_I +h {\bf P}_I^h\right]
       ~~~~,\eqno(24)$$
where $\left(\ddsigma\right)_0=(\sigma_M/M_d) F_0$
is the cross section with unpolarized beam and target. Therefore,
we have the following
relation between the ${\bf p}'s$ and ${\bf P}'s$
$$\eqalign{F_0 {\bf P}_I &= {\bf p}_I ~~~~,\cr
           F_0 {\bf P}_I^h &= {\bf p}_I^h ~~~~.\cr}
           \eqno(25)$$
\par
Since the outgoing nucleon polarization depends on six independent
vectors, one has
in general, as many as 18 different components to study.
They halve in the in-plane kinematics
if the deuteron symmetry axis ${\bf d}$ lies in the reaction plane,
case to which we confine ourselves with the choice $\theta_d=90^\circ$,
$\phi_d=0^\circ$.
In this case
only the component of
$\pz$, $\pd$, and $\ppru$
normal to the reaction plane and
the two components of
$\bppr_0$, $\pu$, and $\pprd$
lying in the reaction plane
are non vanishing
because of parity conservation. In fact, the electron helicity $h$
and the vector orientation parameter $\pdeu$ change sign
while the tensor orientation parameter $\pded$ remains unchanged
under space inversion.\par
In principle the separation of the different polarization vectors
in Eq.(24) from the unavoidable $\pz$
is a simple task, dictated as it is by the range of values covered
by $h$,\ $\pdeu$, \ and $\pded$.
We recall that $h=\pm 1$ for relativistic electrons and that
${\sqrt {2\over3}}\pdeu = {n_1}-{n_{-1}}$
and ${\sqrt 2}\pded = 1-3{n_0}$, where $n_m$ is the fraction of the
deuterons with spin projection $m$ on the symmetry axis ${\bf d}$.
Therefore,
${\sqrt {2\over3}}\pdeu$
varies in the range $(-1,1)$
and ${\sqrt 2}\pded$ in the range $(-2,1)$.
Clearly, to determine $\bppr_0$
one has to perform two measurements of $\bf P$ with unpolarized target
and with opposite values of the electron helicity and to subtract the results.
Obviously such a double experiment is necessary in the general case but not
in the in-plane kinematics where $\pz$ and $\bppr_0$ are
orthogonal, as noted above.
Similarly, $\pu$ comes from the difference of $\bf P$ measured with
unpolarized beam and vector polarized deuteron target with $\pm \pdeu$.
To obtain $\ppru$ one needs four experiments with the four combinations
of $\pm h$ and $\pm \pdeu$ and the appropriate sums and differences
of the resulting polarizations.
The same procedure allows to
disentangle $\pd$ and $\pprd$
because also ${\sqrt 2}\pded$ can assume opposite values in the range $(-1,1)$.
\par
Much more complicated is the problem of disentangling
the different PSF defining the polarization vectors
(see Eq.(12)).
It usually requires several measurements in different
kinematic conditions, at different
values of polarization parameters and also exploiting the
dependence of the polarization components on the polar and
azimuthal angles of ${\bf d}$.
The similar problem
concerning the various SF of the cross section
has been discussed in Ref.[34].
Since the determination of PSF is beyond the present
experimental feasibility we concentrate our discussion directly
on the polarization vectors.
\par
Finally, we address the question of the transformation to the $lab$ frame
of cross section and nucleon polarization. In order to have the differential
cross section fully in the $lab$ frame it is sufficient to multiply
Eq.(18) by the Jacobian
$${\partial \Omega^{cm}_p \over \partial \Omega^{lab}_N} =
  {1\over\gamma} { \left[ \gamma^2 \left( \beta/\beta^{cm}~
                                         + ~cos~\vartheta^{cm} \right)^2
                          + sin^2\vartheta^{cm} \right]^{3/2} \over
                  1 + (\beta/\beta^{cm}) cos~\vartheta^{cm}  }
                 ~~~~,\eqno(26)$$
because the differentials of the electron energy and solid angle in Eq.(18)
are already $lab$ variables. In Eq.(26) $\beta^{cm}$ is the $cm$ nucleon
velocity, $\beta$ and $\gamma$ are the boost parameters.
\par
The transformation of the nucleon polarization is less trivial because the
nucleon momentum is in general not colinear with the boost. Therefore,
the polarization vector undergoes a Wigner rotation [38]
about the axis
${\bf q}\times{\bf p}_{cm}$ through an angle $\psi$ which can be put
in the form
$$tan~\psi = { sin~\vartheta^{cm} \over \gamma^{cm}
               \left(\beta^{cm}/\beta~ + ~cos~\vartheta^{cm} \right) }
           ~~~~.\eqno(27)$$
Hence, the boost leaves unchanged the $y$ component of the polarization
but mixes the $x$ and $z$ components. Explicitly, one has:
$$\eqalign{ P^{lab}_s &= P_x cos~\psi + P_z sin~\psi ~~~, \cr
            P^{lab}_l &=-P_x sin~\psi + P_z cos~\psi ~~~, \cr}
            \eqno(28)$$
where $P^{lab}_l$ and $P^{lab}_s$ are the longitudinal and sideways $lab$
components in the $\phi=0$ half-plane. Note that there is no mixing for
forward and backward emitted nucleons.
\finepagina
\centerline{\bf III. RESULTS}
\vskip 15pt
We shall consider the observables in three theoretical approximations,
the IA theory which corresponds to the standard NR calculations
with nucleonic only contributions to the reaction amplitude and
final state interaction (FSI) included;
the IA+MEC+IC theory which takes into account the mesonic
exchange currents and the $\Delta$-excitation current of pionic
range, the last one in the static approximation for the $\Delta$
propagator; the full theory which also includes the nucleonic
relativistic corrections to the EM operators as well as to the
wave functions and the pionic RC to the charge density in PV
coupling theory.
In actual calculations we use the Paris potential, as said in the
Introduction, and for the nucleon EM form factor we
use the dipole fit and the Galster model [39] of $G_E^n$
(with p=5.6). The sensitivity to $G_E^n$ models is investigated
by also considering the same dipole fit and $G_E^n=0$.
\par
For more details about meson exchange and relativistic corrections
we refer to Ref.[12]
where the technical problem caused by the low
convergence of the multipole expansion of the transition amplitude
for the nucleonic contributions
is also discussed. The short range of the mesonic operators ensures
a good convergence of the mesonic parts of the $t$-matrix.
As a result, we explicitly evaluate Coulomb,
electric, and magnetic multipoles up to order $L=6$ with FSI
while all the higher multipoles are calculated in PW approximation.
This is obtained evaluating the PW amplitude in closed form, i.e.
without recourse to the multipole expansion,
as first shown by Renard et al. [40] and widely discussed
by Fabian and Arenh\"ovel [41]. In order to avoid double
counting one has to subtract the   first
multipoles calculated in PW approximation.
To be fully consistent in such a procedure, we have
chosen the
Foldy gauge [42] (see also Friar and Fallieros [43])
in all calculations (closed form in PW approximation and
multipole expansion without and with FSI).
We would like  to note that the use of the Siegert
form of the transverse electric multipoles allows us to also
account for the major part of the pionic RC through the knowledge
of the pionic charge density.
\par
For simplicity, we do not report results obtained in PW
approximation for the final states.  We recall that in such an
approximation the reduced transition matrix is real in
compliance with Watson's final state theorem. As a consequence,
several components of the polarization
vectors and notably $\pz$, vanish when  FSI are switched off
because they depend on  the imaginary part of appropriate
$p$-functions (see Eqs.~(13-14)). The  inadequacy of this
approximation in describing the angular shape of the
other components, which are given by the real part of
PSF, has been already demostrated in Ref.[22-24]. A noticeable
exception which is worth mentioning, is the
forward direction.
Here, as shown by Arenh\"ovel et al. [22], these polarization components
are essentially given by PWIA.
\par
As kinematical region we have chosen the quasi-free scattering (defined
by the Bjorken variable $x=1$ ) at $Q^2 = 12\fm-2$ and then for
relative n-p energy of about 120 MeV, a condition which maximizes the
momentum transfer involved and therefore the importance of RC
while remaining below the pion production threshold where the description
of the NN interaction in terms of the realistic NN potentials is valid.
To completely fix the kinematics we assume the electron scattering angle
${\vartheta_e}=60^\circ$. The corresponding electron beam energy is
of 820 MeV and the scattered electron energy of 570 MeV, namely a
situation well within the possibilities of the existing accelerators.
For the polarization vector we take  a reference frame slightly
different from that in Sect.~2. More precisely, we leave unchanged
the z-axis in the direction of the nucleon momentum, but we take
the y-axis parallel to ${\bf k}\times{{\bf k}^\prime}$, $\bf k$ and
${\bf k}^\prime$ being the initial and final electron momenta.
Had we followed the Madison convention of Sect.~2. with the y-axis
directed along ${\bf q}\times{\bf p}_{cm}$, the y-components would
have shown an unpleasant
change of sign for nucleon emission to the right or left of $\bf q$.
Polarizations are drawn as a function of the nucleon $cm$ polar
angle measured from the virtual photon direction. Therefore,
$\vartheta_{cm} = 0^\circ$ is the strict quasi elastic peak. The sign
of $\vartheta_{cm}$ has been assumed positive in
the half-plane $\phi=0^\circ$
and negative in the half-plane $\phi=180^\circ$.
\par
We start considering the nucleon polarization in the
unpolarized case. As already said, only the
component of $\pz$ normal to the reaction plane survives
in the in-plane kinematics because of parity conservation.
The first remark worth doing is that the nucleon polarization takes
on appreciable values which make it accessible for measurements.
\par
The proton polarization ${\apz0y}(p)$
reported in Fig.~1b shows a
remarkable sensitivity to both MEC and RC for all angles except
a narrow forward region. In particular these effects add
coherently leading to a reduction by a factor 2 of the maximum
centered at $\vartheta_{cm} =-50^\circ$ and to the disappearence of the peak
at $\vartheta_{cm} =70^\circ$.
Also the pronounced maximum at $\vartheta_{cm} =150^\circ$ suffers from
a $15\%$ lowering. At higher angles in the half-plane $\phi=180^\circ$
MEC and RC interfere destructively resulting in a small and
oscillating polarization.
\par
The neutron polarization
${\apz0y}(n)$ is drawn in Fig.~1a. It is characterized in IA theory by
a sharp maximum followed by a bump  and a deep minimum
in the left half-plane  and by a large maximum with a peak at
$\vartheta_{cm}=130^\circ$ in the right half-plane. While the
minimum is slightly modified by MEC and RC, the shape of the two maxima is
heavily affected. In particular, the bump at $\vartheta_{cm}=-110^\circ$
and the peak at $\vartheta_{cm}=130^\circ$ reduce to two shoulders
owing to the coherent MEC and RC effect.
\par
These results are for calculations with $F_1$. If $G_E$ is used instead,
the angular behaviour of the nucleon
polarization remains essentially the same with its rich structure
of maxima and minima. However, the absolute values show sizeable differences
both in the IA predictions which are smaller on the average and in the
IA+MEC+IC theory. With inclusion of RC one gets  full theory
curves very similar to those in Fig.~1.
\par
If the electron beam is polarized and the target unpolarized, the nucleon
polarization vector acquires a new contribution, the beam polarization
transfer $\bppr_0$, which is
parallel to the reaction plane in the in-plane kinematics.
The higher values of the nucleon polarization are reached by
the longitudinal component $\ppr_{0z}$ in the forward direction
as shown in Fig.~2. Two other maxima appear at backward
angles in the two half planes. Both protons and neutrons present
a positive  polarization transfer, i.e. they are emitted with spin aligned
with the direction of motion.
The IA theory fix the shape of $P_{0z}^h(n)$ in the whole
angular range while the IA+MEC+IC  results are close to
the full theory values.
Indeed, the relativistic effect is visible in the shift towards
the right of the large forward peak and in backward direction.
Instead, RC play a very important role in defining the longitudinal
polarization of the proton either in the forward peak (with a
$30\%$ increase at $\vartheta_{cm}=0^\circ$) and in the
half-plane $\phi =180^\circ$.
As for the calculations with $G_E$ we would like to remark
that the relativistic effect is weaker but still noticeable
in $P_{0z}^h(p)$, particularly
in the backward peaks and at $0^\circ$ where it amounts
to a $10 \%$ increase.
\par
The sideways component $P_{0x}^h$ of $\bppr_0$ shows in general  less
pronounced
but still measurable values with the noticeable exception of the deep
minimum at $\vartheta_{cm}=30^\circ$ for emitted protons, Fig.~3b.
Such minimum is largely due to
the nucleonic NR contributions. Mesonic and relativistic corrections
are effective
at higher angles where however, ${\ppr_{0x}}(p)$ assumes rather small values.
Instead, RC considerably modify $\ppr_{0x}$ for emitted neutrons, Fig.~3a,
in the
forward region with an upward shift of the polarization which even
changes sign in parallel kinematics. We recall
that a measurement of ${\ppr_{0x}}(n)$ has been suggested by Arnold,
Carlson and Gross [18]
for an experimental determination of $G_E^n$ because of its
sensitivity to the different models of $G_E^n$.
The additional curve in Fig.~3a , corresponding
to calculation with ${G_E^n}=0$, sets the scale of such sensitivity.
The dependence of ${\ppr_{0x}}(n)$ on
potential models and on exchange effects has been shown to be
very small at $\vartheta_{cm}=0^\circ$ [21,22].
We can add that it is also little sensitive to the choice of the
Dirac or Sachs form of the nucleon EM form factors. Indeed,
the full theory results with $G_E$ [14] are almost identical
to those in Fig.~3.
We may also note that here the RC effect is generally negligible and that
it almost completely vanishes in parallel kinematics in $P_{0x}^h(n)$.
Then the IA+MEC predictions for $P_{0x}^h(n)$ of Ref.[21-22] obtained with
the Sachs parametrization almost coincide with our full theory results
around $\vartheta_{cm}=0^\circ$. In order to compare our results with those
of Ref.[21-22] one has to recall that Arenh\"ovel et al. report neutron
polarization as function of the proton polar angle and for neutron
emitted in the  $\phi =180^\circ$ half-plane. Moreover, there is a
difference of sign due to our convention for the $x$ component.
\par
In the case of unpolarized beam and polarized target, the nucleon
polarization takes contribution from two new vectors,
(in short, vector and tensor target polarization transfer) $\pu$ and $\pd$
which correspond to the possible vector and tensor deuteron polarization.
Such polarization vectors have not been studied till now in the literature,
probably because their measurement seemed quite unlikely.
As discussed in the Introduction, this is no
longer the case because of the continuous improvements in nucleon
polarimeters and in the production of polarized internal target which,
together with the availability of continuous wave electron beams,
make it possible to foresee experiments of these observables in the
near future. For this reason we report
their non-vanishing components (recall that
$\pu$ lies in the reaction plane while
$\pd$ is normal to that plane
in the coplanar kinematics)
for each possible detected nucleon. The more interesting components
of the target polarization transfers
are the longitudinal component for outgoing neutron and the
sideways component for outgoing proton in the case of vector
polarized deuterons.
In particular ${\bpu1z}(n)$, Fig.~4a,
seems as promising as ${\ppr_{0x}}(n)$
for an experimental determination of $G_E^n$
even if the measurement of a longitudinal polarization presents
an additional difficulty. Indeed, it must be turned
into a transverse polarization
%
%
to be measured with the focal plane polarimeters.
In fact, ${\bpu1z}(n)$ shows the same sensitivity of
${\ppr_{0x}}(n)$
to $G_E^n$ in the forward direction
as the comparison between the full theory curves with
$G_E^n = 0$ and $G_E^n \not =0$ clearly demonstrates.
This is easily understood looking at the analytic expression for
${\bpu1z}(n)$ in PWIA and neglecting the D-wave component of the deuteron
state. In this limit ${\bpu1z}(n)$ which depends in general
on four PSF (see Eq.~(12)), is determined at $0^\circ$ only by the
longitudinal-transverse PSF, i.e. by the interference between charge
and magnetic neutron form factors. We want to note that our
results for   $G_E^n = 0$ do not vanish, as one could naively expect,
not only because of the influence of MEC and FSI and the D-wave deuteron
component but also because our calculations are with the
Dirac form of nucleon current (obviously $F_1 \not = 0$
even for $G_E=0$).
Note that similarly to
${\ppr_{0x}}(n)$,
${\bpu1z}(n)$
is considerably
affected by RC in the forward region where the inclusion of RC causes
a change of sign. The higher
values of ${\bpu1z}(n)$ are attained at higher angles in both
half-planes,
where RC are still of
importance.
In particular, the relativistic effect completely cancels the
exchange effect in the maximum at $\vartheta_{cm}=60^\circ$,
where the sensitivity
to $G_E^n$ is still considerable.
Also in this observable full theory calculations with $G_E$
give very similar results even if the relativistic effect is
much less pronounced.
The results corresponding to the Sachs form factors as reported in
Fig.~5 , show that the large difference at $\vartheta_{cm}=0^\circ$
between calculations with $G_E^n = 0$ and  $G_E^n \not = 0$
does not depend on the nucleon current parametrization.
Here, however, the  $G_E^n = 0$ curve is almost vanishing, so
demonstrating the small influence of FSI and MEC and of the
D-deuteron state at this point. From Fig.~5  it also clearly
appears that RC are less effective in calculations with $G_E$.
A few remarks are worth doing concerning  the different role played
by mesonic and relativistic corrections in the two calculations.
With reference to the peak centered around $\vartheta_{cm}=0^\circ$,
we may note that an almost identical final result is obtained
through a large negative MEC effect balanced by a large positive
RC effect in calculations with
$F_1$. Instead, in calculations with $G_E$, the higher IA
values are considerably lowered by MEC and very slightly
increased by RC. Also remarkable is the fact that the relativistic
effect may even change sign in the two cases, as is evident
in the large peak centered at $\vartheta_{cm}=-100^\circ$.
Mesonic and relativistic effects are small  in   ${\bpu1z}(p)$
in both calculations with $F_1$ (Fig.~4b) and $G_E$ (Fig.~5b)
except in the region from $\vartheta_{cm}=-70^\circ$
to  $\vartheta_{cm}=-140^\circ$
where MEC+IC contribute significantly.
\par
The sideways proton polarization ${\apu1x}(p)$, which is reported
in Fig.~6b, shows considerable variations due to RC over a large
range of $\vartheta_{cm}$.
In fact, a  rather strong relativistic effect is present not only
in the forward region but also
in the two deep minima at medium angles.
It is worth mentioning that even if not
so marked the RC effect remains visible also in $G_E$ calculations.
Unlike  ${\bpu1z}(n)$,  ${\apu1x}(n)$ (Fig.~6a) is little sensitive
to $G_E^n$ and also little affected by MEC and RC.
\par
The normal component of the tensor target polarization transfer
$P_{2y}$ is reported in Fig.~7.
For both proton
and neutron it is characterized by positive and negative peaks
where its magnitude reaches values close to 1. Such values diminish
in the minima and increase in the maximum when RC are considered
as a consequence of a general upward shift induced by such corrections.
The exchange effect is instead very small except in  $P_{2y}(p)$
around
$\vartheta_{cm}=-100^\circ$. Also the dependence on $G_E^n$ is quite
negligible.
\par
It still remains to be considered the case where both beam and target
are polarized. The recoil nucleon polarization takes on two new
contributions denoted $\ppru$ and $\pprd$ in Eq.(24) which can be called
vector and tensor target - beam polarization transfers.
Since the measurement
of nucleon polarization in such conditions is certainly beyond
the range of the present experimental possibilities, we shall only
briefly comment on the results of our calculations. $\ppru$, which is
normal to the reaction plane, assumes very small values in the whole
angular range for both proton and neutron emission, and is
little sensitive to exchange and relativistic corrections (Fig.~8).
Instead,
$\pprd$ which lies in the reaction plane, can reach appreciable values
in the longitudinal (Fig.~9) and sideways (Fig.~10) components.
Whereas ${\apprd2x}(n)$ is mainly determined by IA
results, ${\apprd2x}(p)$ presents interesting features at backward
angles, where it shows appreciable sensitivity to RC and
especially to $G_E^n$ models as revealed by the additional curve
corresponding to full theory calculations with $G_E^n = 0$.
This sensitivity of ${\apprd2x}(p)$ on $G_E^n$ has the same origin
as that previously found for $P_{1z}(n)$. Indeed, a PWIA calculation
with the dominant S-wave deuteron component shows that ${\apprd2x}(p)$
at $180^\circ$ depends linearly on the charge form
factor of the neutron.
Unfortunately, in the quasi elastic peak region, the protons
emitted at $180^\circ$ in the $cm$ frame move in forward direction
in the $lab$ frame with very small kinetic energy. Consequently,
a measurement of their polarization is extremely difficult.
\par
Finally,  ${\cpprd2z}$ has a characteristic structure with four
deep minima where the exchange effect is more pronounced, with
particular evidence in the neutron case. Also RC affect
${\cpprd2z}$ in the minima at $\vartheta_{cm}=-150^\circ$ and
$\vartheta_{cm}=40^\circ$.
Once again the relativistic effect considerably diminishes in
the Sachs scheme.
\finepagina
\centerline{\bf IV. CONCLUSIONS}
\vskip 15pt
We have considered outgoing nucleon polarization
in the case of beam and target polarized, giving the general
formulas for the components, valid in any kinematic conditions. In the
numerical applications, we have studied all the non-vanishing components of
proton and neutron polarization
in the coplanar kinematics and in the quasi-elastic region
assuming the deuteron symmetry axis in the reaction plane.
\par
Our calculations have been done in the standard theory with
inclusion of the most relevant relativistic corrections.
We have found that such corrections give substantial modifications
in almost every polarization component. This does not come out as a
surprise because we have used the Dirac form of the nucleon
current in our calculations. As previous works on coincidence
disintegration indicate, the size of the relativistic effect is
normally larger with this form than with the Sachs form  but the
full theory results are very close to each other. Having verified that
this is indeed the case in all the observables presented,
we have chosen to report only one set of theoretical results
with the exception of $P_{1z}$.
As noted in discussing  $\bpu1z(n)$, the similarity of full
theory results turns out from a complicated interplay of the
IA, MEC and RC contributions which are seldom individually
different in size and even in sign.  The fact that the predictions
are little dependent on the parametrization of the nucleon
current greatly reduces the ambiguity in the theoretical analysis
of deuteron electrodisintegration at the quasi elastic peak.
\par
In particular, we have studied the dependence of nucleon
polarization on the neutron charge form factor, comparing results obtained
with $G_E^n = 0$ and with the dipole form factor model of
Galster et al. [39]. Apart from reobtaining the well known
results of the sensitivity of  $P_{0x}^h(n)$ on $G_E^n$
in parallel kinematics we have picked out two new observables
as sensitive as  $P_{0x}^h(n)$  on $G_E^n$ models.
The first one is the longitudinal component of the vector target
polarization transfer, again in forward direction. A measurement of
$\bpu1z(n)$ seems to be well within the range of the experimental
possibilities in the very near future in view of the recent
advances in the polarized target technique and in the
focal plane polarimeters.
The second observable is the sideways component of tensor target - beam
polarization
transfer for emitted protons  at $\vartheta_{cm}=180^\circ$ which
constitutes instead a prohibitively difficult experiment.
\finepagina
{\bf Acknowledgements}\finerigo
This work was partly supported by Ministero della Universit\`a e
della Ricerca Scientifica of Italy.
\finepagina
{\bf References}\finerigo
\smallskip
\item{[1]~~}  H. Hyuga and M. Gari, Nucl. Phys. {\bf A274}, 333 (1976).
\item{[2]~~}  J. L. Friar , Ann. Phys.(NY) {\bf 104}, 380 (1977).
\item{[3]~~}  W. Jaus and W. S. Woolcock, Helv. Phys. Acta {\bf57}, 644 (1984).
\item{[4]~~}  J. Adam Jr., E. Truhlik and D. Adamova,
              Nucl. Phys. {\bf A492},556 (1991).
\item{[5]~~}  H. G\"oller and H. Arenh\"ovel,
              Few-Body Systems {\bf 13}, 117 (1992).
\item{[6]~~}  K. Tamura, T. Niwa, T. Sato and H. Ohtsubo,
              Nucl. Phys. {\bf A 536},597 (1992).
\item{[7]~~}  J.-F. Mathiot , Phys. Rep. {\bf 173},63 (1989).
\item{[8]~~}  A. Cambi, B. Mosconi and P. Ricci:
              a) Phys. Rev. Lett. {\bf 48} (1982) 462;
\item{\phantom{[8]~~}}b) Proc. Workshop on Perspectives in Nuclear Physics
                 at Intermediate Energies, Trieste, 1983, ed. S.Boffi,
                 C. Ciofi degli Atti and M. M. Giannini
                 (World Scientific, Singapore, 1984) pag.139;
              c) J. Phys. {\bf G10} (1984) L11.
\item{[9]~~}  R.J. Hughes, A. Zieger, H. W\"affler and B. Ziegler,
              Nucl. Phys. {\bf A267},329 (1976).
\item{[10]~~} B. Mosconi and P. Ricci, Few-Body Systems, Suppl. {\bf 5},
              36 (1992).
\item{[11]~~} H. Arenh\"ovel and M. Sanzone, Few-Body Systems,
              Suppl. {\bf 3}, 1 (1991).
\item{[12]~~} B. Mosconi and P. Ricci, Nucl. Phys. {\bf A 517}, 483 (1990).
\item{[13]~~} M. van der Schaar, H. Arenh\"ovel, H.P. Blok,
              H. J. Bulten, E. Hummel, E. Jans, L. Lapik\'as,
              G. van der Steenhoven, J.A. Tjon, J. Wesseling and
              P.K.A. de Witt Huberts,
              Phys. Rev. Lett. {\bf 68}, 776 (1992).
\item{[14]~~} B. Mosconi, J. Pauschenwein and P. Ricci,
              Few-Body Systems, Suppl. {\bf 6}, 223 (1992).
\item{[15]~~} E. Hummel and J.A. Tjon, Phys. Rev. Lett. {\bf 63}, 1788 (1989);
              Phys. Rev. C {\bf 42}, 423 (1990).
\item{[16]~~} C. Giusti and F.D. Pacati, Nucl. Phys. {\bf A 336}, 427 (1980).
\item{[17]~~} R. Madey et al., Proceed. AIP Conference, {\it ``Intersections
              between particle and nuclear physics''} (Tucson, May, 1991)
              edited by W.T.H. Van Oers, pag.~954.
\item{[18]~~} R. G. Arnold, C. E. Carlson and F. Gross,
              Phys. Rev. C {\bf 23}, 363 (1981).
\item{[19]~~} R. Madey et al., Research Proposal 85-05 to MIT-Bates
                               Research Program (1985).
\item{[20]~~} M.P. Rekalo, G.I. Gakh and A.P. Rekalo,
                           Ukr. Phys. J. {\bf 31}, 1293 (1986).
\item{[21]~~} H. Arenh\"ovel, Phys. Lett. B{\bf 199}, 13 (1987).
\item{[22]~~} H. Arenh\"ovel, W. Leidemann and E.L. Tomusiak,
              Z. Phys. {\bf A 331}, 123 (1988);
              Erratum, Z. Phys. {\bf A334}, 363 (1989).
\item{[23]~~} A.Yu. Korchin, Yu.P. Mel'nik and A.V. Shebeko,
              Sov.J. Nucl. Phys. {\bf 48},243 (1988);
              Few-Body Systems {\bf 9}, 211 (1990).
\item{[24]~~} Yu. P. Mel'nik and A. V. Shebeko,
              Few-Body Systems {\bf 13}, 59 (1992).
\item{[25]~~} J.F.J. van den Brand, Nucl. Phys. {\bf A~546}, 299c (1992).
\item{[26]~~} R. Gilman, R.J. Holt, E.R. Kinney, R.S. Kowalczyk,
              S.I. Mishnev, J. Napolitano, D.M. Nikolenko, S.G. Popov,
              D.H. Potterveld, I.A. Rachek, A.B. Temnykh,
              D.K. Toporkov, E.P. Tsentalovich, B.B. Wojtsekhowski
              and L. Young, Phys. Rev. Lett. {\bf 65},1733 (1990).
\item{[27]~~} B. Frois, Nucl. Phys. {\bf A~527}, 357c (1991).
\item{[28]~~} R.J. Holt, Proceed. AIP Conference, {\it ``Intersections
              between particle and nuclear physics''} (Tucson, May, 1991)
              edited by W.T.H. Van Oers, pag.~175.
\item{[29]~~} F. Gross and D.O. Riska, Phys. Rev. {\bf C36}, 1928 (1987).
\item{[30]~~} K. Ohta, Phys. Rev. C {\bf 40}, 1335 (1989).
\item{[31]~~} F. Gross and H. Henning, Nucl. Phys. {\bf A537}, 344 (1992).
\item{[32]~~} H.W.L. Naus and J.H. Koch, Phys. Rev.~C {\bf 39}, 1907 (1989).

\item{[33]~~} W. Leidemann, E.L. Tomusiak and H. Arenh\"ovel,
              Phys. Rev. C {\bf 43}, 1022 (1991).
\item{[34]~~} H. Arenh\"ovel, W. Leidemann and E.L. Tomusiak,
              Phys. Rev. C {\bf 46}, 455 (1992).
\item{[35]~~} G. Beck and H. Arenh\"ovel, Few-Body Systems
              {\bf 13}, 165 (1992).
\item{[36]~~} M. Lacombe , B. Loiseau , J.M. Richard , R. Vin Mau,
              J. C\^ot\'e, J. Pir\'es, and R. De Tourreil,
              Phys. Rev. C {\bf 21}, 861 (1980).
\item{[37]~~} V. Dmitrasinovic and F. Gross, Phys. Rev. C{\bf 40},2479 (1989);
              Erratum , Phys. Rev. C {\bf 43},1495 (1991).
\item{[38]~~} A.D. Martin and T.D. Spearman, ``{\it Elementary particle
theory}''
              (North - Holland, Amsterdam, 1970).
\item{[39]~~} S. Galster, H. Klein, J. Moritz, K.H. Schmidt, D. Wegener
              and J. Bleckwenn,
              Nucl. Phys. {\bf B 32}, 221 (1971).
\item{[40]~~} R.M. Renard, J. Tran Thanh Van and M. Le Bellac,
              Nuovo Cim. {\bf 38} (1965), 565; 1688.
\item{[41]~~} W. Fabian and H. Arenh\"ovel,
              Nucl. Phys. {\bf A 314}, 253 (1979).
\item{[42]~~} L.L. Foldy, Phys. Rev. {\bf 92}, 178 (1953).
\item{[43]~~} J.L. Friar and S. Fallieros, Phys. Rev. C {\bf 29}, 1645 (1984);
              Phys. Rev. C {\bf 34}, 2029 (1986).
\finepagina
\parindent=20pt\parskip=12pt plus 1pt
\leftline{\bf FIGURE CAPTIONS}
\bigskip
\item{\bf Fig.~1~~} Outgoing neutron (a) and proton (b) polarization $P_{0y}$
                    as a function of the corresponding polar angle.
                    Dot-dashed line : IA theory; dotted line: IA + MEC + IC
                    theory; solid line: full theory.
                    Calculations with the Paris potential,
                    with the Dirac form of nucleon current and
                    with the dipole Galster model of EM form factors.

\item{\bf Fig.~2~~} Beam polarization transfer $P^h_{0z}$.
                    Notation as in Fig.1.

\item{\bf Fig.~3~~} Beam polarization transfer $P^h_{0x}$.
                    Notation as in Fig.1. The additional curve (dashed
                    line) corresponds to $G_E^n=0$.

\item{\bf Fig.~4~~} Vector target polarization transfer $P_{1z}$.
                    Notation as in Fig.3.

\item{\bf Fig.~5~~} The same as in Fig.4 with the Sachs
                    form of nuclear current.

\item{\bf Fig.~6~~} Vector target polarization transfer $P_{1x}$.
                    Notation as in Fig.1.

\item{\bf Fig.~7~~} Tensor target polarization transfer $P_{2y}$.
                    Notation as in Fig.1.

\item{\bf Fig.~8~~} Vector target - beam polarization transfer $P_{1y}^h$.
                    Notation as in Fig.1.

\item{\bf Fig.~9~~} Tensor target - beam polarization transfer $P_{2z}^h$.
                    Notation as in Fig.1.

\item{\bf Fig.~10~~}Tensor target - beam polarization transfer $P_{2x}^h$.
                    Notation as in Fig.3.
\bye